%% file: paper.tex
\parskip = 0pt 
\def\makeheadline{\vbox to 0pt{\vskip-30pt\line{\vbox to8.5pt{}\the
\headline}\vss}\nointerlineskip}

\input stbasic.tex

\twelvepoint

\doublespace

\raggedbottom
\null\vskip.5in
\vskip.4in
\centerline{\bf The discovery of a giant debris arc in the
Coma Cluster}
\smallskip
\medskip
\medskip
\centerline{Neil Trentham}
\vskip 4pt
\centerline{Institute for Astronomy, University of Hawaii}
\vskip 4pt
\centerline{2680 Woodlawn Drive, Honolulu HI 96822, U.~S.~A.}
\vskip 4pt
\centerline{and}
\vskip 4pt
\centerline{Bahram Mobasher}
\vskip 4pt
\centerline{Astrophysics Group}
\vskip 4pt
\centerline{Blackett Laboratory, Imperial College}
\vskip 4pt
\centerline{Prince Consort Road, London SW7 2BZ, U.~K.}
\vskip.3in
\vskip.3in
\centerline{Submitted to {\it MNRAS}} 
\vskip.3in
\vskip.3in
\vfil
\eject
 
\centerline{\bf ABSTRACT }
\bigskip

\noindent
We present the discovery a giant low surface-brightness arc, of length
$\sim 80$ kpc in the Coma Cluster.  The arc consists of a diffuse
luminous matrix with surface-brightness $\mu_B < 26.5$ mag arcsec$^{-2}$
and a number of embedded condensations.
It is not associated with any giant galaxy in Coma in particular; 
neither does it have the properties of a gravitational arc.
We argue that a fast interaction between the nearby barred S0 galaxy
IC 4026 and either IC 4041 or RB 110 is the most natural explanation
for the origin of the arc. 


\vfil\eject

\noindent{\bf 1 INTRODUCTION}

\noindent
Interactions between galaxies are an important mechanism driving
galaxy evolution.  
One aspect of this process that has been investigated in detail is
the interactions between galaxies in clusters.  
This is because the galaxy densities in clusters is high, and so 
observations lead to a robust statistical description 
of the galaxy population there, at least at bright magnitudes 
(see e.g.~Binggeli 1994).   

Present-day clusters have high observed velocity dispersions
($\sim 1000$ km s$^{-1}$), close to the virial values.  
This means that interactions
between galaxies in clusters normally occur at high relative
encounter velocities, 
which in turn means that the energy transfer between
the two galaxies is small (Binney \& Tremaine 1987).  One might
therefore expect violent interactions between galaxies 
like those described by Toomre \& Toomre (1972) to be
rare in clusters.    

However, observations tell us that galaxies in clusters {\bf are}
undergoing significant evolutionary changes.  In particular, the
blue Butcher-Oemler (1978, 1984) galaxies seen in clusters at redshifts
$z>0.2$, are not seen at $z=0$.  A number of explanations of this
have been hypothesized (e.g.~Lavery \& Henry 1988, Kauffmann 1995),
based on the notion that  
interactions may indeed play an important role if they occur during the
early stages of cluster formation before
virialization, when the galaxy velocities are small.  More
recently,  
Moore et al.~(1996) have investigated the effects of galaxy
harassment  
i.e.~the cumulative effect 
large numbers of high velocity  
encounters in the cluster gravitational potential well. 
Their models show that harassment can simulatenously account for
disappearance of the Butcher-Oemler
galaxies 
and the properties of dwarf elliptical
galaxies seen in large number in low-redshift clusters.

So interactions may indeed be important in shaping galaxies in clusters.
We might therefore expect to see features like low surface-brightness
tidal tails and merger debris (Barnes \& Hernquist 1992)
in clusters, particularly in young clusters that are not dynamically
relaxed, and in clusters which are merging with other clusters.
Indeed, Moore et al.~(1996) describe the existence of 
giant debris arcs in
clusters as a natural consequence of their harassment models.

We now report the discovery  
of such an arc or tail, of length $\sim 80$ kpc,
in the Coma cluster ($z=0.023$), located 13.4 arcminutes
from the cluster center.
Such features were difficult to find previously because of their
very low surface brightnesses and the large angular sizes of nearby
clusters like Coma, and the consequent need to survey   
large areas at depth.  The new large-format CCDs that are used
on large telescopes now permit such surveys
to be made, and the arc we describe here was independently discovered in
two such surveys.  
It is perhaps not surprising that such a feature should be found in
Coma, where the absence of a cooling flow (Stewart et al.~1984) and
the existence of two large central galaxies of approximately equal
optical luminosity are suggestive that Coma might be the result of
a recent merger between two smaller clusters.  If this is true, it
will not be a relaxed system today.

\vskip 10pt

\noindent{\bf 2 OBSERVATIONS} 

\noindent
The arc (see Fig.~1)
was discovered in a deep two-color ($B$ and $R$) survey
of Coma (Trentham 1997), conducted using a thinned 
backside-illuminated Tektronix
2048 X 2048 CCD at the f/10 Cassegrain focus of the University of
Hawaii 2.24 m telescope on Mauna Kea. 
The exposure times for this
survey were 35 minutes in the $B-$band and 25 minutes
in $R-$band, using Mould filters, with a median
seeing of 0.75$^{\prime \prime}$
FWHM.  The arc was
confirmed independently in the data from a second survey, using the 4.2 m
William Herschel Telescope on La Plama. It has very low 
surface-brightness ($< 26.5$  mag
arcsec$^{-2}$ in the $B-$band), but is clearly visible above the
sky background (see Figure 1c).  The center of the arc
is at ($\alpha$(1950),$\delta$(1950)) =
(12$^{\rm h}$58$^{\rm m}$5.3$^{\rm s}$,
28$^{\circ}$19$^{\prime}$52.52$^{\prime \prime}$).
The arc is about 3$^{\prime}$ long, has a radius of
approximately 40$^{\prime \prime}$,
and is located at about 13.4$^{\prime}$ from the cD galaxy NGC 4874 in
the direction N62.9$^{\circ}$E and 7.05$^{\prime}$
from the other supergiant Coma cluster galaxy
NGC 4889 in the
direction N43.2$^{\circ}$E (at the distance of the Coma cluster one
arcminute corresponds to 26 kpc).  No radio or X-ray source is known at the
location of the arc.

Figure 1c shows that the arc consists of a diffuse luminous matrix and a
number of embedded condensations.  The arc looks approximately circular with
radius 40$^{\prime \prime}$,
but in fact is poorly fit by a circle or ellipse; the mean residual
from the best-fit circle to the five brightest condensations that are not
background galaxies is about 10$^{\prime \prime}$.
The photometric properties of the arc are
shown in Figure 2.  The brightest objects in the arc are labelled, and their
colours and magnitudes are presented in Table 1.
The relative height of the $B$ and $R$
lines in Figure 2 gives a good estimate of the colour of each object.  Red
objects are mostly background; a giant elliptical galaxy in Coma has
$B-R = 1.9$
(Coleman et al.~1980)
so objects redder than this are probably background galaxies.

In Figure 2, we also display the surface-brightness detection thresholds
required to find the arc.  Only the more recent
CCD surveys approach the faint limits required to detect the arc.
The 1$\sigma$ sky noise of the UH 2.24 m data was 28.3 
$B$ mag arcsec$^{-2}$.
The total area of the UH 2.24 m survey was 674 arcminutes, and this was the
only such feature found $-$ such objects are presumably either very
short-lived or very rare, even in clusters like Coma that we 
suspect to be unrelaxed.
Previous photographic surveys of Coma (e.g.~Thompson \& Gregory 1993) 
routinely surveyed areas much larger than 
this, but were not deep enough to find
features like this arc.  Previous deep CCD surveys 
(e.g.~Bernstein et al.~1995) easily reached sky noise levels this low,
but did not cover sufficient area to find these features either. 
It is the combination of area and depth that allow the new
large-format CCD surveys like the UH 2.24 m and the Herschel 4.2 m surveys
to detect objects such as this. 

The arc may continue beyond the limits shown in Figure 2, but its
surface-brightness
would then have to be less than the sky noise
in this region.   Figure 2 also shows that the luminous
matrix is visible above the sky
over the entire length of the arc;
while this
may in part be due to the way we have geometrically 
defined the arc, Figure 1c
confirms that there is indeed a substantial amount of flux seen between the
condensations.  The maximum surface brightness of the matrix is 26.5 mag
arcsec$^{-2}$ in the $B-$band; this corresponds to object P.

The arc is not gravitational, as suggested by the following observations.
Firstly, the arc is resolved in the radial direction, having a thickness
of up to 10$^{\prime \prime}$,
about ten times the seeing.  Gravitational arcs, on the other
hand, are normally radially unresolved.  
Secondly, many of the embedded objects are
resolved and do not have a colour distribution consistent with them all being
part of a single background galaxy.   Finally, for an Einstein ring with a lens
at the distance of Coma,  the angular radius
$\theta$ must satisfy the inequality $\theta <
9.6 \left({ {M}\over{10^{11}{\rm M}_{\odot}}}\right)^{0.5}$ arcseconds,
where $M$ is the enclosed mass.  The observed
angular radius of about 40$^{\prime \prime}$
would then require an enclosed mass of at
least $2 \times 10^{12}{\rm M}_{\odot}$.
This is unreasonably large for the region enclosed by the arc, which
is 13.4$^{\prime}$  from the cluster center,
and in which there is no optical emission.  In
any case, the Einstein ring interpretation is probably incorrect, as the arc is
so poorly fit by a circle.

\vskip 10pt

\noindent{\bf 3 DISCUSSION} 

\noindent
The arc is similar in appearance to the tidal features described by
Barnes \& Hernquist (1992); the condensations therein might then be
dwarf galaxies in formation.  However, unlike the tails in 
the simulations, the arc here is not clearly associated with any
giant galaxy in particular.  Table 2 lists the nearby galaxies; their
locations can be seen in Figure 1(b).  If the arc is debris from an
interaction between galaxies, the galaxies involved must therefore have
traveled some distance from the arc. 

The large distances between the arc and the nearest galaxies suggest that
if the arc results from interaction, the encounter velocities of the 
galaxies involved must be high. 
The galaxy harassment simulations of Moore et al.~(1996) may 
therefore offer some 
additional clues to the origin of the arc. 
If the harassment models provide a reasonable prescription for these
observations i.e.~the arc is debris from a fast encounter between two
galaxies, we ought to be able to identify both a harassed and
harassing galaxy.
We refer to the galaxy providing the material for the
arc as the harassed galaxy, and the one with which it has experienced
a recent fast encounter as the harassing galaxy.
The present-day 3-D velocity dispersion of Coma is
1010 km s$^{-1}$  (Zabuloff et al.~1990)
so that the 1-D velocity dispersion in the plane
of the sky is $\sim 0.3$ degree/Gyr = 20 arcmin/Gyr
($H_0  = 75$ km s$^{-1}$ kpc$^{-1}$).
Therefore one might expect the harassed
galaxy to be within a few arcminutes of the arc.
The barred S0 galaxy, IC 4026
is far and
away the closest giant galaxy in Coma on the concave side of the arc, and so is
the best candidate for being
the harassed galaxy.  It is 2.0$^{\prime}$ from the center of
curvature of the arc in the direction S60$^{\circ}$E.
It is an attractive candidate for the harassed galaxy
because fast encounters can create bars in the
harassed galaxy (Moore et al.~1996).  Furthermore, IC 4026 has a
similar color to most of the objects in the arc, as is predicted by the
harassment models.  A few objects (e.g.~E and I) have slightly bluer colours;
this is probably due to small amounts of recent star formation in the debris.
If this interpretation is correct, then the time since harassment is
approximately 0.1 $\left({ {V_{ah}} \over {500 \,
{\rm km s}^{-1}}}\right)^{-1}$
Gyr, where $V_{ah}$  is the present-epoch
relative velocity of the arc and IC 4026 in the plane of the sky.   The most
natural place to search for the harassing galaxy is  the opposite side of the
arc from IC 4026 in a direction opposite to S60$^{\circ}$E i.e.
N60$^{\circ}$W.  Candidate
galaxies include IC 4041 (5.1$^{\prime}$
from IC 4026 in the direction N53$^{\circ}$W) and 
RB 110  (S0 galaxy, 4.2$^{\prime}$ from IC 4026 in the
direction N62$^{\circ}$W).
From Table 2, the projected component of the encounter
velocity perpendicular to the plane of the
sky is $\sim 1100$ km/s for IC 4041 and $\sim 700$ km/s for RB 110;
these numbers are consistent with a fast encounter
interpretation. 
For these galaxies, the time since
harassment is approximately
0.5 $\left({{d} \over {10^{\prime}}}\right)
\left({{V_{gh}} \over {500 \, {\rm km s}^{-1}}}\right)^{-1}$
Gyr, where $V_{gh}$
is the present-epoch relative velocity of the harassing galaxy and IC 4026 in
the plane of the sky and $d$ is the angular separation of the two galaxies, as
stated above.  
For a value
of $V_{gh}$ of twice the 1-D velocity dispersion of Coma, the time
since harassment would be 0.22 Gyr and 0.18 Gyr for IC 4041 and RB 110,
respectively.  
If the encounter velocity is higher (Moore et al.~consider encounter
velocities as high as 1500 km s$^{-1}$), these times would be
shorter, and the arc would be younger. 
While we have argued, mostly on {\it a posteriori} grounds,
that a fast interaction is the likeliest explanation for the
arc, it should also be recgnized that lower velocity
hyperbolic interactions could also produce such a feature.
Given the high observed velocity dispersion of Coma, it is
perhaps unlikely that such an interaction would occur between
cluster members (again on {\it a posteriori} grounds).
But the arc, if produced in a low-velocity encounter, would be
dynamically cold, and the parent galaxies could
then lie well away from it in the plane of the sky, further than
any of the galaxies listed in Table 2.  

Comparison between this arc and either the tails simulated by
Barnes \& Hernquist, or the arcs simulated by Moore et al.~therefore
does reveal some similarities.  The main 
difference in both cases seems to be that the arc is not clearly
associated with any galaxy in particular, although the timescales
involved suggest that the fast encounter hypothesis is plausible.
As more simulations are carried out, in particular with differing
harassed galaxy types, encounter velocities and geometries, and
local environments,
it will hopefully become clear whether this feature of the arc that
we see can be readily explained.  
What determines the position of the center of curvature of the arc
(it is coincident with neither IC 4026 or the cluster center)
is another problem that is not readily addressed by the existing
simulations. 
Measuring the arc redshift would be another a useful test of our 
interpretation, as well as a possible means of identifying the
harassed and harassing
galaxes (the arc redshift should be close to, preferably intermediate
between, the redshifts of the harased and harassing galaxies).  However,
it is
not at present feasible to do this, even with the largest telescopes,
as the arc
is too faint.
Redshifts of the brightest two or three objects within the arc
may be obtainable, but we cannot be sure that these are at the same redshift as
the diffuse material.

The colours of the objects in the arc (see Table 1), suggest very little
ongoing star formation there,
as most objects are red, with $B-R$ of 1.4
to 1.6 (similar to IC 4026).
This suggests that most of the material in the arc
is stellar debris from the harassed galaxy, and that there is probably very
little gas in the arc; this is perhaps not surprising as the
harassed galaxy is
a S0 galaxy --
these are normally gas-poor.  Objects C and H are very blue, and
may be globular clusters in formation.  However, they are very faint ($M_B = -
10$), and will fade even more as they get older.

Throughout this discussion we have
assumed that the diffuse material in the arc lies in the
Coma Cluster.
This is not rigorously proven
(only a spectroscopic redshift for the diffuse material would
do that), and we assess the alternative possibilites here.
Some of the condensations (like A, B, and D) are probably
background galaxies, as they have colours typical of
old stellar populations at $z \approx 0.2$.  
The arc is probably not associated with
these because the colour of the diffuse material is significantly
bluer than the colours of these galaxies (this might be
caused by ongoing star formation, but it would then have to be
spread over a feature 0.8 Mpc in size).
A more serious
potential worry is that the arc is in fact a
foreground object that happens to lie in front of the
Coma Cluster.  Then its size would not be as big as we claim.
However, no object similar to this has been seen in any of the
very deep wide-field surveys of the field and nearby groups
of galaxies undertaken by the authors
(ten times the area of this survey have
been surveyed down to $R = 26.5$ mag
arcsec$^2$).
This, combined with the concordance between the colours of the arc
and many Coma Cluster galaxies, suggests that the
most straightforward interpretation of the arc at this stage is
that it belongs to the Coma Cluster.

To summarize, we have discovered a giant debris arc 
in Coma and regard fast encounters between
nearby galaxies as the likeliest explanation of its properties, although
the comparison between the observed properties of the arc and existing
simulations raise a number of new questions. 
New simulations with input parameters derived from the observed parameters of
the galaxies identified here
will also be useful in further assessing the nature of the arc. 

\vskip 10pt

\noindent{\bf ACKNOWLEDGMENTS}

\noindent
This research has made use of the NASA/IPAC extragalactic database (NED) which
is operated by the Jet Propulsion Laboratory, Caltech, under agreement with the
National Aeronautics and Space Administration.

\vfill\eject

\ni{\bf REFERENCES }

\beginrefs

\noindent
Barnes J.~E., Hernquist L., 1992, Nat, 360, 715

\noindent
Bernstein G.~M., Nichol R.~C., Tyson J.~A., Ulmer M.~P., Wittman D., 1995, AJ,
110, 1507

\noindent 
Binggeli B., 1994, in Meylan G., Prugneil P., ed., ESO
Conference and Workshop Proceedings No.~49: Dwarf Galaxies. 
European Space Observatory, Munich, p.~13

\noindent
Binney J., Tremaine S., 1987, Galactic Dynamics, Princeton University Press,
Princeton 

\noindent
Butcher H., Oemler A., 1978, ApJ, 219, 18

\noindent
Butcher H., Oemler A., 1984, ApJ, 285, 426  

\noindent 
Coleman G.~D., Wu C.-C., Weedman D.~W., 1980, ApJS, 43, 393

\noindent
Kauffmann G., 1995, MNRAS, 274, 161

\noindent
Lavery R.~J., Henry J.~P., 1988, ApJ, 330, 596

\noindent
Moore B., Katz N., Lake G., Dressler A., Oemler A., 1996, Nat, 
379, 613

\noindent
Rood H.~J., Baum W.~A., 1967, AJ, 72, 398  

\noindent
Stewart G.~C., Fabian A.~C., Jones C., Forman W., 1984, 
ApJ, 285, 1

\noindent
Thompson L.~A., \& Gregory S.~A., 1993, AJ, 106, 2197

\noindent 
Toomre A., Toomre J., 1972, ApJ, 178, 623

\noindent
Trentham N., 1997, MNRAS, submitted 

\noindent
Zabuloff A.~L., Huchra J.~P., Geller M.~J., 1990, ApJS, 74, 1

\endrefs
\vfil
\eject
 
\noindent {\bf FIGURE CAPTIONS}

\vskip 10pt

\noindent {\bf Figure 1.~(a)} 
Region of the Coma cluster surveyed by Trentham (1997); the survey
covered $\sim 97$\% of the area shown here.  This
image comes from the Palomar Sky Survey.  It 
is 26 arcminutes square.  The box indicates the location of the smaller
area displayed in (b).  Here, as in all three images, North is up
and East is to the left; {\bf (b)} 
A 6.5 arcminute square region containing the arc, and showing clearly
the giant galaxies in the vicinity of the arc.  This image is from our
$B$-band UH 2.24 m data.  The six large galaxies in the frame,
clockwise from the upper right are IC 4026, RB 100, IC 4042, IC 4041,
RB 110, and IC 4040.  All are S0 galaxies, except IC 4040, which is a
very disturbed Sdm galaxy.  The box shows the location of the smaller
area displayed in (c); {\bf (c)} The $B$-band image of the arc, from our
UH 2.24 m data.  The image appears somewhat granular; this contrast is
best to show the low-surface-brightness arc against the sky background.

\vskip 10pt

\noindent {\bf Figure 2.~(a)} The variation
in $B-$band  and $R-$band flux density within a 3.0$^{\prime \prime}$
diameter circular aperture, with position along the arc.  The photometry
presented here is from our sky-subtracted UH 2.24 m data.  
The seeing was 0.75 arcseconds in $B$ and 0.76 arcseconds in $R$.
The distance along
the arc (the ordinate parameter) was computed as follows.  The
n$^{\rm th}$ point P$_n$ was
found by calculating the vector {\bf r} joining the n-2$^{\rm th}$
point P$_{n-2}$ to the n-1$^{\rm th}$
point P$_{n-1}$.
We then add {\bf r} to the position vector of P$_{n-1}$ to get the point
(P$_n$)$_0$.  The line joining P$_{n-1}$
to (P$_n$)$_0$ was then rotated through $\pm 45^{\circ}$ about P$_{n-1}$
to define a circular segment of arc S$_n$
centered on (P$_n$)$_0$.  The total flux
contained in a 3.0$^{\prime \prime}$
diameter circular aperture was then computed for each point
along S$_n$, and the point P$_n$
chosen to be the point at which this flux is
maximized.  If we then define a distance between points
(set to be 4.5$^{\prime \prime}$;
this is
small enough to give an acceptable resolution), and two initial points (set to
be the points at $\pm 2.25^{\prime \prime}$
due north of object N, where the arc is running
north-south), the arc is then uniquely defined.  Positive directions are
clockwise, and the minimum and maximum
distances on the ordinate axes correspond to the ends of the arc i.e.~the
places where the flux within a 3.0$^{\prime \prime}$ diameter
aperture drops to zero.
The thick horizontal lines indicate average surface-brightnesses within the
aperture  
of 24, 25, 26, 27, 28 mag arcsec$^{-2}$, from top to bottom in the $B$ (left)
and $R$ (right) bands;  
{\bf (b)} The location on the arc of the condensations described in the
text and labelled in (a).  The contours are from our $B$-band image
(Figure 1(c)) and cover a 110 arcsecond square region with North up
and East to the left.  Some of the redder condensations appear only very
faintly on this $B$-band image.  

\par\vfill\eject\bye

%% file: stbasic.tex
\message{STBASIC.TEX TeX Macro Library}
\message{ }





\def\beginrefs{\begingroup\parindent=0pt\frenchspacing
   \parskip=1pt plus 1pt minus 1pt\interlinepenalty=1000\pretolerance=10000
   \hyphenpenalty=10000\everypar={\hangindent=0.42in}       
  \def\aa##1{{\it Astr.~Ap., \bf ##1}}
  \def\aasup##1{{\it Astr.~Ap.~Suppl., \bf ##1}}
  \def\aasupp##1{{\it Astr.~Ap.~Suppl., \bf ##1}}
  \def\aj##1{{\it A.~J., \bf ##1}}
  \def\annrev##1{{\it Ann.~Rev.\ Astr.~Ap., \bf ##1}}     
  \def\araa##1{{\it Ann.~Rev.\ Astr.~Ap., \bf ##1}}     
  \def\apj##1{{\it Ap.~J., \bf ##1}}     
  \def\apjl##1{{\it Ap.~J. (Letters), \bf ##1}}
  \def\apjlett##1{{\it Ap.~J. (Letters), \bf ##1}}
  \def\apjlet##1{{\it Ap.~J. (Letters), \bf ##1}}
  \def\apjsup##1{{\it Ap.~J.~Suppl., \bf ##1}}
  \def\apjsupp##1{{\it Ap.~J.~Suppl., \bf ##1}}
  \def\baas##1{{\it Bull.~A.A.S., \bf ##1}}
  \def\ban##1{{\it B.A.N., \bf ##1}}
  \def\ibvs##1{{\it Inf. Bull. Var. Stars}, No.~##1}
  \def\mn##1{{\it M.N.R.A.S., \bf ##1}}
  \def\mnras##1{{\it M.N.R.A.S., \bf ##1}}
  \def\pasp##1{{\it Pub.~A.S.P., \bf ##1}}
  \def\ajpasp##1{{\it Pub.~A.S.P., \bf ##1}}
  \def\nat##1{{\it Nature, \bf ##1}}
  \def\nature##1{{\it Nature, \bf ##1}}}

\def\endrefs{\endgroup}



\def\df{\leaders\hbox to 0.6em{\hss.}\hfill}


\def\section#1{\bigbreak\medskip\centerline{#1}\par\nobreak\medskip\markpage}

\def\subsection#1#2{\bigbreak\noindent{\bf#1\hskip 0.9em\relax#2}\par
   \nobreak\medskip\markpage}

\def\subsubsection#1#2{\medbreak\noindent{\sl#1\hskip 0.60em\relax#2}\par
   \nobreak\medskip\markpage}

\def\today{\advance\year by -1900 
   \number\month/\number\day/\number\year}
\def\yearmonthday{\number\year\space
   \ifcase\month\or January\or February\or March\or April\or May\or June\or
   July\or August\or September\or October\or November\or December\fi
   \space\number\day}

\newcount\num

\def\nextnum{\global\advance \num by 1 \number\num}
\def\nextitem{\leavevmode
   \hbox{\ifnum\num>8 \kern-0.43em\fi \nextnum.\kern0.60em}}
\def\bfnextitem{\leavevmode
   \hbox{\ifnum\num>8 \kern-0.43em\fi \bf\nextnum.\kern0.60em}}

\newcount\colnum

\def\nextcolnum{\global\advance \colnum by 1 \number\colnum}
\def\nextcolumn{\leavevmode
   \hbox{{\it \ifnum\colnum<9 \phantom{1}\fi Column \nextcolnum:}\kern0.60em}}

\newcount\fig

\def\nextfig{\global\advance \fig by 1 \number\fig}

\newcount\cap

\def\nextcap{\global\advance \cap by 1 \number\cap}

\newcount\letter

\def\nextlet{\global\advance \letter by 1
   \ifcase\letter\or A\or B\or C\or D\or E\or F\or G\or H\or I\or
   J\or K\or L\or M\or N\or O\or P\or Q\or R\or S\or T\or U\or V\or W\or X\or
   Y\or Z\fi}

\newdimen\bigindent \bigindent=3.5in
\def\letterhead{\hsize=6in\interlinepenalty=2000\parskip=6pt minus 3pt
  \pretolerance=750
  \def\topline##1{\hbox to\hsize{\hfil##1\hskip\rightskip}}
  \footline={\ifnum\pageno=1
    \hss\hbox{\vrule height 0.4in width 0pt}
    \eightrm Operated by the Association of Universities for Research in 
    Astronomy, Inc., for the National Aeronautics and Space Administration\hss
    \else\hfil\fi}
  \null
  \vskip-0.2in
  {\advance\rightskip by -0.75in
    \topline{3700 San Martin Drive}
    \topline{Baltimore, MD 21218}
    \topline{(301) 338-4718}\par}
  \vskip30pt minus 15pt
  {\leftskip=\bigindent\yearmonthday\par}}

\def\arpanetletterhead{\hsize=6in\interlinepenalty=2000\parskip=6pt minus 3pt
  \pretolerance=750
  \def\topline##1{\hbox to\hsize{\hfil##1\hskip\rightskip}}
  \footline={\ifnum\pageno=1
    \hss\hbox{\vrule height 0.4in width 0pt}
    \eightrm Operated by the Association of Universities for Research in 
    Astronomy, Inc., for the National Aeronautics and Space Administration\hss
    \else\hfil\fi}
  \null
  \vskip-0.2in\vskip-3\baselineskip
  {\advance\rightskip by -0.75in
    \topline{3700 San Martin Drive}
    \topline{Baltimore, MD 21218}
    \topline{(301) 338-4718}
    \topline{{\elevenrm BITNET:} \tt golombek@stsci}
    \topline{\elevenrm SPAN: \tt SCIVAX::GOLOMBEK}
    \topline{{\elevenrm ARPANET:} \tt golombek@scivax.arpa}\par}
  \vskip30pt minus 15pt
  {\leftskip=\bigindent\yearmonthday\par}}

\def\gosbletterhead{\hsize=6in\interlinepenalty=2000\parskip=6pt minus 3pt
  \pretolerance=750
  \def\topline##1{\hbox to\hsize{\hfil##1\hskip\rightskip}}
  \footline={\ifnum\pageno=1
    \hss\hbox{\vrule height 0.4in width 0pt}
    \eightrm Operated by the Association of Universities for Research in 
    Astronomy, Inc., for the National Aeronautics and Space Administration\hss
    \else\hfil\fi}
  \null
  \vskip-0.375in
  {\advance\rightskip by -0.75in
    \topline{General Observer Support Branch}
    \topline{3700 San Martin Drive}
    \topline{Baltimore, MD 21218}
    \topline{(301) 338-4996}\par}
  \vskip30pt minus 15pt
  {\leftskip=\bigindent\yearmonthday\par}}



\def\indentleft{\advance\leftskip by 50pt\interlinepenalty=750}
\def\inndentleft{\advance\leftskip by 78pt\interlinepenalty=750}
\def\narrower{\advance\leftskip by 0.42in\advance\rightskip by 0.42in
  \interlinepenalty=750}
\def\nnarrower{\advance\leftskip by 50pt\advance\rightskip by 45pt
  \interlinepenalty=750}

\def\checkbox{\nnarrower\parindent=0pt\itemitem{\vbox{\hrule height.7pt
  \hbox{\vrule width.7pt height6pt \kern6pt \vrule width.7pt}
  \hrule height.7pt}$\,$}}  


%
%
\newcount\index \index=100
\def\markpage{\advance\index by 1 \count\index=\pageno}
\def\begintableofcontents{\begingroup
  \index=100 \frenchspacing\interlinepenalty=750
  \parskip=0.1pt plus 1pt minus 0.1pt \parindent=0.3in
  \def\dfi{\advance\index by 1 \df\number\count\index}
  \def\in{\par\hskip-0.2in\indent \hangindent2\parindent \textindent}    
  \def\inin{\par\hskip0.32in\indent \hangindent3\parindent \textindent}
  \def\ininin{\par\hskip0.95in\indent \hangindent4\parindent \textindent}}



{\obeylines\gdef\startdisplay#1
  {\catcode`\^^M=5$$#1\halign\bgroup\indent##\hfil&&\qquad##\hfil\cr}}
\outer\def\enddisplay{\crcr\egroup$$}

\chardef\other=12

{\obeyspaces\gdef {\ }} 

  \font\twentyfourrm=cmr10 scaled 2488
  \font\twentyfouri=cmmi10 scaled 2074   
  \font\twentyfoursy=cmsy10 scaled 2074
  \font\twentyrm=cmr10 scaled 2074      
  \font\twentyi=cmmi10 scaled 2074   
  \font\twentysy=cmsy10 scaled 2074
  \font\eighteenrm=cmr10 scaled 1728
  \font\eighteeni=cmmi10 scaled 1728 \font\eighteensy=cmsy10 scaled 1728
  \font\fourteenrm=cmr10 scaled 1440
  \font\fourteeni=cmmi10 scaled 1440 \font\fourteensy=cmsy10 scaled 1440
  \font\twelverm=cmr12
                
  \font\twelvei=cmmi12               \font\twelvesy=cmsy10 scaled 1200
  \font\elevenrm=cmr10 scaled 1095
    
  \font\eleveni=cmmi10 scaled 1095   \font\elevensy=cmsy10 scaled 1095
  \font\tenrm=cmr10
                   
  \font\teni=cmmi10  \font\tensy=cmsy10  
  \font\ninerm=cmr9

  \font\ninei=cmmi9                  \font\ninesy=cmsy9
  \font\eightrm=cmr8
  \font\seveni=cmmi7 \font\sevensy=cmsy7

\def\commonstuff{
  \parindent=0.42in       
  \def\skipline{\vskip\baselineskip}
  \hyphenpenalty=200\pretolerance=300\tolerance=600 
  \interlinepenalty=100\clubpenalty=500\widowpenalty=500
  \nonfrenchspacing\singlespace\rm}

\def\twelvepoint{
  \font\bf=cmbx12
  \font\it=cmti12
  \font\sl=cmsl12
  \font\tb=cmtt10 scaled 1200 
  \font\tt=cmtt8 scaled 1440
  \textfont0=\twelverm \scriptfont0=\tenrm     
    \scriptscriptfont0=\sevenrm                 
  \def\rm{\fam0 \twelverm}   
  \textfont1=\twelvei  \scriptfont1=\teni  
    \scriptscriptfont1=\seveni                  
  \def\mit{\fam1 } \def\oldstyle{\fam1 \twelvei}
  \textfont2=\twelvesy \scriptfont2=\tensy 
    \scriptscriptfont2=\sevensy                 
  \def\singlespace{\baselineskip=13.5pt\lineskiplimit=-5pt
    \lineskip=0pt
    \parskip=1.25pt plus 1.5pt minus 0.25pt}  
  \def\oneandahalfspace{\baselineskip=18pt\parskip=0pt plus 1pt}
  \def\doublespace{\baselineskip=24pt\parskip=0pt plus 0.5pt}
  \footline={\ifnum\pageno=1 \hfil
             \else\hss\twelverm-- \folio\ --\hss\fi} 
  \def\pagenumbers{\footline={\hss\twelverm-- \folio\ --\hss}}  
  \def\romanpagenumbers{\footline={\hss\twelverm-- \romannumeral\folio\ --\hss}}
  \commonstuff}

\def\tenpoint{
  \font\it=cmti10
  \font\sl=cmsl10
  \font\bf=cmb10
  \textfont0=\tenrm \scriptfont0=\sevenrm     
    \scriptscriptfont0=\fiverm                 
  \def\rm{\fam0 \tenrm}   
  \textfont1=\teni  \scriptfont1=\seveni  
    \scriptscriptfont1=\fivei                  
  \def\mit{\fam1 } \def\oldstyle{\fam1 \teni}
  \textfont2=\tensy \scriptfont2=\sevensy 
    \scriptscriptfont2=\fivesy                 
  \def\singlespace{\baselineskip=12pt\lineskiplimit=0pt
    \lineskip=-0.5mm       
    \parskip=2pt plus 1pt minus 1pt}  
  \footline={\ifnum\pageno=1 \hfil
             \else\hss\tenrm-- \folio\ --\hss\fi} 
  \def\oneandahalfspace{\baselineskip=18pt\parskip=0pt plus 1pt}
  \def\doublespace{\baselineskip=24pt\parskip=0pt plus 1 pt}
  \def\pagenumbers{\footline={\hss\tenrm-- \folio\ --\hss}}  
  \def\romanpagenumbers{\footline={\hss\tenrm-- \romannumeral\folio\ --\hss}}
  \commonstuff}

\def\elevenpoint{
  \font\it=cmti10 scaled 1095
  \font\sl=cmsl10 scaled 1095
  \font\bf=cmb10 scaled 1095 
  \font\tt=cmtt10 scaled 1095
  \textfont0=\elevenrm \scriptfont0=\tenrm     
    \scriptscriptfont0=\ninerm                 
  \def\rm{\fam0 \elevenrm}   
  \textfont1=\eleveni  \scriptfont1=\teni  
    \scriptscriptfont1=\ninei                  
  \def\mit{\fam1 } \def\oldstyle{\fam1 \eleveni}
  \textfont2=\elevensy \scriptfont2=\tensy 
    \scriptscriptfont2=\ninesy                 
  \def\singlespace{\baselineskip=13pt\lineskiplimit=-5pt
    \lineskip=0mm       
    \parskip=2pt plus 1pt minus 1pt}  
  \footline={\ifnum\pageno=1 \hfil
             \else\hss\elevenrm-- \folio\ --\hss\fi} 
  \def\oneandahalfspace{\baselineskip=19pt\parskip=0pt plus 1pt}
  \def\doublespace{\baselineskip=26pt\parskip=0pt plus 1 pt}
  \def\pagenumbers{\footline={\hss\elevenrm-- \folio\ --\hss}}  
  \def\romanpagenumbers{\footline={\hss\tenrm-- \romannumeral\folio\ --\hss}}
  \commonstuff}

\def\eighteenpoint{           
  \font\bf=cmbx10 scaled 1728
  \font\it=cmti10 scaled 1728
  \font\sl=cmsl10 scaled 1728
  \font\tb=cmtt10 scaled 1728
  \font\tt=cmtt10 scaled 1728
  \textfont0=\eighteenrm \scriptfont0=\fourteenrm
    \scriptscriptfont0=\twelverm                 
  \def\rm{\fam0 \eighteenrm}   
  \textfont1=\eighteeni  \scriptfont1=\fourteeni  
    \scriptscriptfont1=\twelvei                  
  \def\mit{\fam1 } \def\oldstyle{\fam1 \eighteeni}
  \textfont2=\eighteensy \scriptfont2=\fourteensy 
    \scriptscriptfont2=\twelvesy                 
  \def\singlespace{\baselineskip=21pt\lineskiplimit=-5pt
    \lineskip=0pt
    \parskip=4pt plus 1pt minus 1pt}  
  \def\oneandahalfspace{\baselineskip=30pt\parskip=0pt plus 1pt}
  \def\doublespace{\baselineskip=40pt\parskip=0pt plus 1pt}
  \footline={\ifnum\pageno=1 \hfil
             \else\hss\eighteenrm-- \folio\ --\hss\fi} 
  \def\pagenumbers{\footline={\hss\eighteenrm-- \folio\ --\hss}}  
  \commonstuff}

\def\twentypoint{
  \font\bf=cmbx10 scaled 2074
  \font\it=cmti10 scaled 2074
  \font\sl=cmsl10 scaled 2074
  \font\tb=cmtt10 scaled 2074
  \font\tt=cmtt10 scaled 2074
  \textfont0=\twentyrm \scriptfont0=\eighteenrm     
    \scriptscriptfont0=\fourteenrm                 
  \def\rm{\fam0 \twentyrm}   
  \textfont1=\twentyi  \scriptfont1=\eighteeni  
    \scriptscriptfont1=\fourteeni                  
  \def\mit{\fam1 } \def\oldstyle{\fam1 \twentyi}
  \textfont2=\twentysy \scriptfont2=\eighteensy 
    \scriptscriptfont2=\fourteensy                 
  \def\singlespace{\baselineskip=24pt\lineskiplimit=-5pt
    \lineskip=0pt
    \parskip=5pt plus 1.5pt minus 1.5pt}  
  \def\oneandahalfspace{\baselineskip=33pt\parskip=0pt plus 1pt}
  \def\doublespace{\baselineskip=44pt\parskip=0pt plus 0.5pt}
  \footline={\ifnum\pageno=1 \hfil
             \else\hss\twentyrm-- \folio\ --\hss\fi} 
  \def\pagenumbers{\footline={\hss\twentyrm-- \folio\ --\hss}}  
  \def\romanpagenumbers{\footline={\hss\twentyrm-- \romannumeral\folio\ --\hss}}
  \commonstuff}

\def\twentyfourpoint{
  \font\bf=cmbx10 scaled 2488
  \font\it=cmti10 scaled 2488
  \font\sl=cmsl10 scaled 2488
  \font\tb=cmtt10 scaled 2488
  \font\tt=cmtt10 scaled 2488
  \textfont0=\twentyfourrm \scriptfont0=\twentyrm     
    \scriptscriptfont0=\eighteenrm                 
  \def\rm{\fam0 \twentyfourrm}   
  \textfont1=\twentyfouri  \scriptfont1=\twentyi  
    \scriptscriptfont1=\eighteeni                  
  \def\mit{\fam1 } \def\oldstyle{\fam1 \twentyfouri}
  \textfont2=\twentyfoursy \scriptfont2=\twentysy 
    \scriptscriptfont2=\eighteensy                 
  \def\singlespace{\baselineskip=28pt\lineskiplimit=-5pt
    \lineskip=0pt
    \parskip=5pt plus 1.5pt minus 1.5pt}  
  \def\oneandahalfspace{\baselineskip=42pt\parskip=0pt plus 1pt}
  \def\doublespace{\baselineskip=56pt\parskip=0pt plus 0.5pt}
  \footline={\ifnum\pageno=1 \hfil
             \else\hss\twentyfourrm-- \folio\ --\hss\fi} 
  \def\pagenumbers{\footline={\hss\twentyfourrm-- \folio\ --\hss}}  
  \def\romanpagenumbers{\footline={\hss\twentyfourrm-- \romannumeral\folio\ --\hss}}
  \commonstuff}

\def\spose#1{\hbox to 0pt{#1\hss}}
\def\lta{\mathrel{\spose{\lower 3pt\hbox{$\mathchar"218$}}
     \raise 2.0pt\hbox{$\mathchar"13C$}}}
\def\gta{\mathrel{\spose{\lower 3pt\hbox{$\mathchar"218$}}
     \raise 2.0pt\hbox{$\mathchar"13E$}}}

\def\ni{\noindent}
\def\in{\indent}
\def\inin{\in{\in}
\def\ininin{\inin{\in}}}